\documentclass{actasen}

\begin{document}

%%ÉèÖÃÊ×Ò³Ò³Âë
\setcounter{page}{115}

\Volume{2011}{35}% Äê¡¢¾í

%%ҳüÉèÖÃ

\runheading{LAI Xiao-yu}%

\title{Ferromagnetism of electrons in solid quark cluster stars$^{\dag}~ \!^{\star}$}

\footnotetext{$^{\dag}$ Supported by the National Natural Science Foundation of China (11203018), the West Light Foundation (XBBS-2014-23), the Science Project of Universities in Xinjiang (XJEDU2012S02) and the Doctoral Science Foundation of Xinjiang University (BS120107).

Received 2013--05--02; revised version 2013--08--06

$^{\star}$ A translation of {\it Acta Astron. Sin.~}
Vol. 55, No. 2, pp. 116--126, 2014 \\
\hspace*{5mm}$^{\bigtriangleup}$ laixy@pku.edu.cn\\

\noindent 0275-1062/01/\$-see front matter $\copyright$ 2011 Elsevier
Science B. V. All rights reserved. %%

\noindent PII: }

\enauthor{LAI Xiao-yu$^{\bigtriangleup}$ }{School of Physics, Xinjiang University, Urumqi 830046, China\\
Xinjiang Astronomical Observatory, Cinese Academy of Sciences, Urumqi 830011, China}

\abstract{
In this paper we are trying to solve the problem of the origin of strong magnetic fields in the framework of solid quark-cluster stars.
We propose that, under the Coulomb repulsion, the electrons inside the stars could spontaneously magnetized and become ferromagnetic, and hence would contribute non-zero net magnetic momentum to the whole star.
The leading order approximation in our simple model shows that the magnetism of solid quark-cluster matter could be possible.  
For most cases in solid quark-cluster stars, the amount of net magnetic momentum, could be significant, and the net magnetic moments of electron system in solid quark-cluster stars could be large enough to induce the observed  magnetic fields for pulsars with $B\sim 10^{11}$ to $\sim 10^{13}$ Gauss.
This mechanism of generating magnetic field is not valid for so-called magnetars.
}


\maketitle

\section{Introduction}

The states of matter of pulsar-like compact stars is a long-standing problem.
Some efforts have been made to understand the nature of pulsars, among which the model of 
quark-cluster stars has been proposed.
With a stiff equation of state, the model of quark-cluster stars suggests the exist of high mass 
($>2M_\odot$) pulsars~\rf{1,2}, to be favored by the discoveries of massive pulsars.

The origin of strong magnetic fields of pulsar-like compact stars is also a long-standing problem.
Here we try to solve this problem in the framework of solid quark star model.
The magnetic moment of electrons is much larger than that of quark-clusters, so electrons could significantly contribute magnetic moment to the whole star.
If we take the intrinsic magnetic moment of electrons as the possible elements giving rise to the macroscopic magnetism, then it seems to be similar to the case of ferromagnetism of normal material in terrestrial environment.

The ferromagnetism of normal material is studied extensively in condensed matter physics.
The origin of ferromagnetism is often demonstrated by Stoner's model.
It considers screened short-range Coulomb interaction, where
the physical picture of ferromagnetism in repulsive Fermi gases can be 
understood as the result of the competition between the repulsive interaction
and the Pauli exclusion principle.
Electrons tend to have unbalanced spins to save interaction energy, and on the 
other hand they tend to have balanced spins to save kinetic energy.

In fact, in condensed matter physics, a full and complete description of ferromagnetism of electron system is still very complex and has not achieve a satisfying stage.
A microscopic calculation from first principle is certainly very difficult and is not the focus of this paper, but the leading order approximation in our simple model shows that the magnetism of solid quark-cluster matter could be possible.

\section{Electrons in solid quark-cluster stars}

For quark-cluster stars, the situation is in fact more simple because electrons are not confined to ``nuclei''
(i.e. the quark-clusters in lattices), and all of the electrons are itinerant, and a quark-cluster star could serve as
an ideal system where the itinerant ferromagnetism might occur under the Stoner's theory.
In solid quark-cluster stars, the electric charge per baryon $Y_e=n_e/n_b< 10^{-4}$, due to the weak symmetry breaking of light-flavor symmetry, so electrons in solid quark-cluster stars could be seen as dilute Fermi gas.
The individual quark inside each quark-cluster has electric charge, so quark-clusters could be polarized in the presence of electrons, with polarization much larger than that of vacuum.
The screening effect coming from quark-clusters could be significant, which could reduce the long-range Coulomb interaction to short-range screened interaction.
That means, to get significant Coulomb interaction, an electron should be very close to another electron, with distances much smaller than the average distance$ n_e^{-1/3}$.
This make the Stoner's model to be valid, which treats the interaction between electrons as a $\delta-$function with strength proportional to the scattering length of electrons.

Defining the amount of unbalanced spins $\xi=\frac{n_+-n_-}{n_++n_-}=\frac{n_+-n_-}{n_e}$,
where $n_+$ and $n_-$ denote the number density of spin-up and spin-down electrons,
respectively.
We find that, in solid quark-cluster stars, the value of $\xi$ could be non-zero.
In some cases, the corresponding magnetic moment
per unit mass $\mu_0$ could be higher than $\sim 10^{-3}$ Gauss cm$^3$ g$^{-1}$,
which is large enough to induce the observed  magnetic fields of pulsars with $B\sim 10^{11}-10^{13}$ Gauss for a pulsar with $M\sim 1.5 M_\odot$ and $R\sim 10$ km (shown in Fig.1)~\rf{3}.

%%%%%%%%%%%%%%%%%%%%%%%%%%%%%%%%%%%%%%%%
\begin{figure}[tbph]
\centering
{\includegraphics[angle=0,width=9cm]{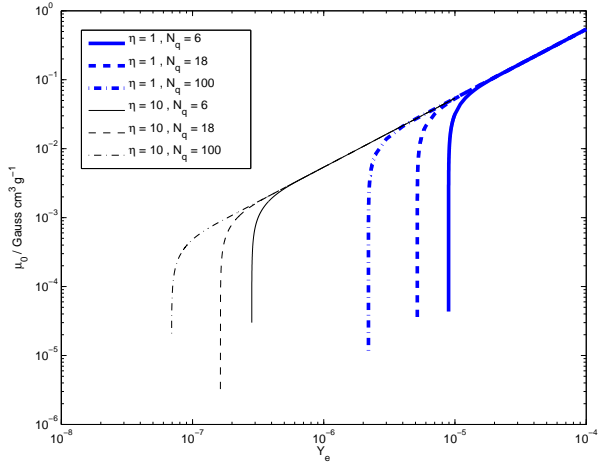}}
%\vspace{-2mm}
\caption{%
The relation between $\mu_0$ (magnetic moment per unit mass) and $Y_e$ (electric charge per baryon $n_e/n_b$), in two cases $\eta=1$ (thick blue lines) and $\eta=10$ (thin black lines), where $\eta$ denotes the strength of polarization, defined as the ratio of electrons' scattering length $a$ and the distance between two nearby quark-clusters $d$.
Solid, dashed and dash-dotted lines correspond to $N_q=$6, 18, 100 respectively ($N_q$ denotes the number of quarks inside each quark-cluster).
If $\mu_0\geqslant 10^{-3}\sim 10^{-2}$ Gauss cm$^3$ g$^{-1}$, then $B\geqslant 10^{12}$ Gauss for a pulsar with $M\sim 1.5 M_\odot$ and $R\sim 10$ km.
\label{fig_mu}}
    \end{figure}

%%%%%%%%%%%%%%%%%%%%%%%%%%%%%%%%%%%%%%%%%

\section{Conclusions and discussions}

We demonstrate that the strong magnetic fields of pulsars  
could be originated from the spontaneous magnetization of electrons
in solid quark-cluster star model.

This mechanism is not valid for so-called magnetars, since extra gravitational and elastic energy, rather than magnetic energy, could be released to power anomalous X-ray pulsars and soft gamma-ray repeaters in the solid quark-cluster star model.

It should also be noted that, the properties of relativistic electrons are in fact not
very certain to us.
Although the Fermi energy of electrons could be as large as $\sim 10$ MeV which means the electrons are relativistic particles, we still use the concept of scattering length in non-relativistic scenario.
Therefore, the model we present in this paper for relativistic electron-system
is to some extent an approximation to the electrons in quark-cluster stars.

\acknowledgements{I like to thank the LOC of ``1st QCS conference''. }

\end{document}